\documentclass[prl,twocolumn,superscriptaddress,aps,amsmath,amssymb,nofootinbib]{revtex4}
\usepackage{graphicx}
\usepackage{dcolumn}
\usepackage{bm}
\usepackage{color}

\usepackage[section]{placeins}
\usepackage{graphicx,epsfig}
\usepackage{amsmath}
\usepackage{amsfonts}
\usepackage{braket}
\usepackage{fancyhdr}
\usepackage{hyperref}
\usepackage[utf8]{inputenc}
\usepackage[T1]{fontenc}
\usepackage{amsthm}
\usepackage{hhline}
\usepackage{multirow}
\begin{document}

  \title{
  $\Delta$-admixed neutron stars: spinodal instabilities and dUrca processes}
\author{Adriana R. Raduta}
\affiliation{National Institute for Physics and Nuclear Engineering
  (IFIN-HH),  RO-077125 Bucharest, Romania}
\begin{abstract}
  Within the covariant density functional theory of nuclear matter we build
  equations of state of $\Delta$-admixed compact stars.
  Uncertainties in the interaction of $\Delta(1232)$ resonance states with nuclear matter,
  due to lack of experimental data, are accounted for by varying the coupling
  constants to scalar and vector mesonic fields. We find that,
  over a wide range of the parameter space allowed by nuclear physics experiments and
  astrophysical observations,
  cold catalyzed star matter exhibits a first order phase transition
  which persists also at finite temperature and out of
  $\beta$-equilibrium in the neutrino-transparent matter.
  Compact stars featuring such a phase transition in the outer core
  have small radii and, implicitly, tidal deformabilities.
  The parameter space is identified where simultaneously $\Delta$-admixed compact stars
  obey the astrophysical constraint on maximum mass and allow for dUrca processes,
  which is otherwise forbidden.
\end{abstract}

\maketitle

\section{Introduction}
\label{sec:intro}

With densities exceeding several times the nuclear saturation density,
$n_s \approx 0.16~{\rm fm}^{-3}$, the core of neutron stars (NS)
has been acknowledged since long ago to represent a unique environment
for studies of compressed baryonic matter.
In the innermost shells of massive stars several non-nucleonic degrees 
of freedom (d.o.f.) - such as hyperons, kaon and pion condensates,
$\Delta$-resonances and quark gluon plasma - have been conjectured
\cite{Glend2000} to exist in addition or instead of
the nucleonic ones.

The high-precision mass measurements, during this decade, of
several massive pulsars in binary systems with white dwarfs \cite{Demorest_2010,Antoniadis2013,Fonseca_2016,Arzoumanian_2018,Cromartie2019}
re-opened the issue of dense matter hyperonisation.
A large variety of theoretical frameworks has been employed, such as
relativistic density functional theory (DFT), 
quark-meson coupling model,
auxiliary field diffusion Monte Carlo approach,
cluster variational method and Brueckner-Hartree-Fock theory.
Within the most frequently employed DFT class of models reconciliation among
two solar mass NS and hyperonic d.o.f. has been possible
either employing a sufficiently stiff nucleonic equation of state (EoS)
or going beyond the SU(6) symmetry ansatz to fix the vector meson couplings,
for a detailed discussion see \cite{Chatterjee_EPJA_2015}.
The relativistic quark model and DFT have been also used to address nucleation of
$\Delta(1232)$-resonances
in stellar matter, see \cite{Sahoo_PRC_2018,Drago_PRC_2014,Cai_PRC_2015,Zhu_PRC_2016,Kolomeitsev_NPA_2017,Li_PLB_2018,Ribes_2019,Li_ApJL_2019,Raduta_MNRAS_2020}.
With masses lying between those of $\Sigma$ and $\Xi$ hyperons
and an attractive potential in nuclear matter, 
$\Delta$-resonances are expected to be populated based on the
same generic energetic arguments with which hyperonisation is advocated.
Similarly to the appearance of any other new hadronic species
above its production threshold $\Delta$s will soften the EoS and,
thus, modify the mass-radius diagram.
Refs. \cite{Cai_PRC_2015,Zhu_PRC_2016} have shown that nucleation of $\Delta$ in purely
nucleonic NS decreases the maximum mass by up to $\approx 0.5M_{\odot}$ and the radii of intermediate
mass NS by up to $\approx 2$ km. The magnitude of these effects is nevertheless very sensitive to the underlying
nucleonic EoS, $\Delta$s effective mass and strength of $\Delta-N$ interaction.
Ref. \cite{Drago_PRC_2014,Kolomeitsev_NPA_2017,Li_PLB_2018,Ribes_2019,Li_ApJL_2019,Raduta_MNRAS_2020}
have shown that in hypernuclear stars the only significant effect induced by the nucleation of
$\Delta$s consists in the reduction of radii.
High precision determination of NS radii, expected from the
currently operating NICER mission \cite{NICER} and also from
future x-ray observatories like Athena x-ray telescope \cite{Athena}
and eXTP \cite{eXTP}, could thus contribute to shed light on the $\Delta$ d.o.f. in NS and, implicitly,
on $\Delta-N$ interaction. Complementary information can, in principle, be provided also by NS cooling
history, considering that, because of its negative charge, $\Delta^-$ can shift the threshold of
nucleonic dUrca to lower densities and, possibly, open new dUrca processes.
The first effect is particularly interesting as hyperonisation has been shown to not alter
the threshold of nucleonic dUrca \cite{Fortin_PRD_2020}.

A sufficiently attractive $\Delta$N interaction potential is expected
not only to modify NS global properties but also the thermodynamic
stability of dense matter. The occurrence of thermodynamic instabilities
related to the onset of $\Delta$s
has been so far discussed in connection with
hot and dense hadronic matter produced in high-energy
heavy-ion collisions \cite{Lavagno_PRC_2010} and, more recently,
cold hypernuclear stars \cite{Lavagno_2019}.
\cite{Lavagno_2019} shown that such instabilities manifest
over a baryonic density domain which corresponds to the inner core,
$4 n_s \lesssim n_B \lesssim 7 n_s$, for coupling constants spanning
wide domains of values in agreement with
available experimental constraints.
The limited number of EoS discussed in \cite{Lavagno_2019} nevertheless
suggests that when the two solar mass constraint is imposed
$\Delta$-driven instabilities are ruled out.

The first aim of the present work is to investigate whether
$\Delta$-driven instabilities occur also in NS built upon EoS which fulfill
the $2M_{\odot}$ constraint \cite{Antoniadis2013}.
To this aim a systematic investigation of the
parameter space is performed. As \cite{Lavagno_2019} we use
the covariant density functional theory but rely on a different nucleonic EoS
and, for the sake of clarity, disregard hyperonic d.o.f.
Muons are also disregarded.
The second motivation of this study is to identify the parameter space of the
$\Delta-N$ interaction where compact stars that obey the two solar mass limit
allow for dUrca processes. To better address this issue we select a
density-dependent nucleonic EoS which does not allow for $n\to p+e^-+\tilde \nu_e$
and $p+e \to n+\nu_e$ dUrca processes in stable stars.

The paper is organized as follows.
In Section \ref{sec: EoS} we present the EoS models.
In Section \ref{sec:instab} the formalism for the analysis of thermodynamic
instabilities in multi-component systems is revisited.
Properties of $\Delta$-admixed compact star are discussed
in Section \ref{sec:res}.
Finally, the conclusions are drawn in Section \ref{sec:concl}.

\section{Equations of state}
\label{sec: EoS}

The phenomenological EoS considered in our study are obtained
by extending the density-dependent DDME2 \cite{Lalazissis_PRC_2005} nucleonic
parameterisation such as to additionally account for $\Delta(1232)$
resonance states of the baryon $J^{3/2}$ decuplet.
Interaction among different baryonic species is realized
by the exchange of scalar-isoscalar ($\sigma$), vector-isoscalar ($\omega$)
and vector-isovector ($\rho$) mesons.
Coupling constants of $\Delta$s to mesonic fields are expressed in terms
of coupling constants of nucleons to mesonic fields
$x_{m,\Delta}=g_{m,\Delta}/g_{m;N}$, where $m=\sigma, \omega, \rho$;
as common in literature, the assumption on the
same density dependence for $\Delta$- and nucleon-meson couplings is made.

Our choice of the nucleonic DDME2 is motivated by the following factors:
i) the parameters of isospin symmetric nuclear matter around
saturation density are in good agreement with present experimental constraints;
ii) the values of the symmetry energy at saturation $J=32.3$ MeV
as well as its slope $L=51.2$ MeV and curvature $K_{sym}=-87.1$ MeV lie within the domains
deduced from nuclear data and neutron star observations;
for constraints on $L$, see \cite{Lattimer_EPJA_2014,Oertel_RMP_2017};
for constraints on $K_{sym}$, see \cite{Mondal_PRC_2017,dEtivaux2019,Zimmerman_2020};
the relatively low value of $L$ explains the good agreement of the energy per baryon
with {\em ab initio} calculations of low density pure neutron matter
\cite{Gandolfi_PRC_2012,Hebeler_ApJ_2013} (see fig. 12 in \cite{Fortin_PRC_2016});
iii) the maximum mass of purely nucleonic NS, $M^{(N)}_{\rm max}=2.48 M_{\odot}$
\cite{Fortin_PRC_2016}, built upon DDME2 fulfills the $2M_{\odot}$ observational
constraint on the lower limit of maximum masses \cite{Antoniadis2013};
iv) the radius of the canonical $1.4M_{\odot}$ NS, $R_{1.4}=13.3$~km, agrees with
the recent measurements $R(1.44^{+0.15}_{-0.14} M_{\odot})= 13.02^{+1.24}_{-1.06}$~km~\citep{Miller_2019} and
$R(1.34^{+0.15}_{-0.16} M_{\odot})= 12.71^{+1.14}_{-1.19}$ km~\citep{Riley_2019}
of the PSR J0030+0451 millisecond pulsar obtained by the NICER mission,
v) for purely nucleonic NS, the tidal deformability range
$826 \leq \tilde\Lambda \leq 854$,
corresponding to the mass ratio range $0.73 \leq q \leq 1$ of the merger stars,
slightly overshoots the upper limit of the interval
$\tilde \Lambda =300^{+500}_{-190}$ (symmetric 90\% credible interval)
for a low spin prior extracted from the GW170817 event \cite{Abbott_PRX_2019},
vi) the agreement with above cited astrophysical data is maintained upon
inclusion of hyperons \cite{Colucci_PRC_2013,Dalen2014,Fortin_PRC_2016,Li_PLB_2018};
vii) the nucleonic dUrca process is not allowed in stable stars;
this provides us with the perfect framework to assess the impact
the nucleation of $\Delta$s has on chemical composition.

In vacuum $\Delta(1232)$s are broad resonances which decay into nucleons with emission
of a pion. In medium they are considered to be stabilized by the Pauli blocking
of the final nucleon states. Other effects of the medium regard the narrowing
of the quasi-particle width and shift of quasi-particle energy to larger values \cite{Sawyer1972,Ouellette2001}.
Within the DFT and relativistic quark model it was assumed that $\Delta$s preserve their
vacuum masses and have vanishing widths \cite{Sahoo_PRC_2018,Drago_PRC_2014,Cai_PRC_2015,Zhu_PRC_2016,Kolomeitsev_NPA_2017,Li_PLB_2018,Ribes_2019,Li_ApJL_2019}.
In the present work we employ the same hypothesis.

The information about $\Delta$-nucleon interaction is scarce.
Data extracted from pion-nucleus scattering and pion photo-production
\cite{Nakamura_PRC_2010}, electron scattering on nuclei \cite{Koch_NPA_1985}
and electromagnetic excitations of the $\Delta$-baryons \cite{Wehrberger_NPA_1989} have been reviewed
by Drago et al. \cite{Drago_PRC_2014} and Kolomeitsev et al. \cite{Kolomeitsev_NPA_2017} with the aim
of constraining the values of the coupling constants at saturation.
Their conclusions may be summarized as follows:
i) the potential of the $\Delta$s in the nuclear medium is
slightly more attractive than the nucleon potential
$-30~{\rm MeV}+U_{N}^{(N)} \lesssim U_{\Delta}^{(N)} \lesssim U_{N}^{(N)}$;
this translates in values of $x_{\sigma\Delta}$ slightly larger
than 1,
ii) $0 \lesssim x_{\sigma \Delta} - x_{\omega \Delta} \lesssim 0.2$, and
iii) no experimental constraints exist for the value of $x_{\rho \Delta}$.

The still remaining uncertainties are customarily accounted for by allowing
for variation of coupling constants within large domains of values.
Previous works \cite{Drago_PRC_2014,Kolomeitsev_NPA_2017,Li_PLB_2018,Ribes_2019,Li_ApJL_2019}
have considered the ranges $0.8 \leq x_{\sigma\Delta} \leq 1.8$,
$0.6 \leq x_{\omega \Delta} \leq 1.6$ and
$0.5 \leq x_{\rho \Delta} \leq 3$.
We shall hereafter adopt the same strategy and the following domains
$0.9 \leq x_{\sigma\Delta} \leq 1.5$;
$x_{\sigma \Delta}-0.2 \leq x_{\omega \Delta} \leq  x_{\sigma \Delta}+0.2$;
$0.7 \leq x_{\rho \Delta} \leq 1.3$.

The core EoS calculated within DFT is smoothly merged with the crust EoS.
For the latter we employ the Negele and Vautherin \cite{Negele_NPA_1973}
and Haensel-Zdunik-Dobaczewski \cite{Haensel_AA_1989} EoS.

\section{Thermodynamic instabilities}
\label{sec:instab}

Spinodal instabilities manifest as convexity anomalies of thermodynamic
potentials expressed in terms of extensive variables.
Mathematically they are signaled by negative eigenvalues of
the curvature matrix
$C_{i,j}=\partial^2 f \left(\{n_i \}; i=1,...,N \right)/\partial n_i \partial n_j$,
where $i,j=1,...,N$;
$f$ stands for the free energy density;
$n_i$ represents the number density of each conserved species $i$,
whose total number is $N$.
The eigenvectors associated to negative eigenvalues correspond to
the directions in the density space along which density fluctuations are
spontaneously and exponentially amplified in order to achieve phase separation.
Each negative eigenvalue corresponds to a phase transition
whose order parameter is given by the eigenvector.
The maximum number of phase transitions in a multi-component system is equal to the
dimension of the curvature matrix which, in its turn, is equal to the number of
conserved species. 
The curvature matrix analysis has been often employed in mean-field
studies of baryonic matter. For the case of liquid-gas phase transition
taking place at sub-saturation densities,
see \cite{Avancini_PRC_2006,Ducoin_NPA_2007,Ducoin_PRC_2008,Torres_PRC_2016};
for strangeness-induced instabilities in dense strange hadronic matter, see
\cite{Gulminelli_PRC_2012,Gulminelli_PRC_2013,Oertel_JPG_2015,Oertel_EPJA_2016,Torres_PRC_2017}.

Spinodal instabilities of multi-component systems whose direction of phase
separation is dominated by one of the conserved densities, $n_j$,
are revealed by convex intruders in the free energy density Legendre conjugated
with respect to the remaining ($N-1$) chemical potentials,
$\bar f(n_j, \{\mu_i\}_{i=1,..,N;i \neq j})=f-\sum_{i=1,..,N; i \neq j} \mu_i n_i$ or, alternatively,
back-bending behavior of the chemical potential
$\mu_j=\left(\partial \bar f/\partial n_j\right)_{n_i; i=1,...,N; i \neq j}$
as a function of the conjugate species density $n_j$.
For a detailed discussion, see \cite{Ducoin_NPA_2006}.

In the most simple case, that we employ here,
in which muons and hyperons are disregarded
$\Delta$-admixed stellar matter is made of neutrons, protons, the four
$\Delta$-isobars ($\Delta^-$, $\Delta^0$, $\Delta^+$, $\Delta^{++}$)
and electrons.
The net charge neutrality condition links the baryon charge density,
$n_Q=n_p+n_{\Delta^+}+2 n_{\Delta^{++}}-n_{\Delta^-}$,
to the electron density, $n_e=n_Q$.
In the neutrino-transparent matter regime - characterized by vanishing lepton chemical potential,
$\mu_L=0$, - the electron density equals the lepton density, $n_e=n_L$.
This means that the thermodynamic potentials may be expressed
in terms of the total baryon number density, $n_B$, and
charge (or lepton) density, $n_{Q(L)}$.

In the following spinodal instabilities will be identified via
the negative eigenvalues of $C_{B,L}=\partial^2 e(n_B, n_L)/\partial n_B \partial n_L$,
where $e$ stands for the
total energy density, the relevant thermodynamic potential at zero temperature.
Despite the fact that, being interested in neutron stars, we impose
$\beta$-equilibrium, the stability analyses are performed in two dimensions,
which means that we allow density fluctuations to drive matter out of
$\mu_L=0$.
Phase transitions will be discussed in the hybrid ensemble
$(n_B, \mu_L)$~\cite{Ducoin_NPA_2006} for $\mu_L=0$,
which corresponds to $\beta$-equilibrated matter.
Extension to finite temperatures and non-vanishing values of $\mu_L$
will allow us to check the persistence of
$\Delta$-driven instabilities in hot matter out of $\beta$-equilibrium.

\section{Results}
\label{sec:res}

We now turn to investigate chemical composition and thermodynamic instabilities
of cold catalyzed $\Delta$-admixed stellar matter, 
for different values of the coupling constants of $\Delta$ to mesonic fields.
The onset of $\Delta$ isobars is discussed in connection with
density thresholds of dUrca processes involving nucleons,
$n \to p + e^- + \tilde \nu_e$,
as well as the two most abundant $\Delta$s: 
$\Delta^- \to n + e^- + \tilde \nu_e$,
$\Delta^0 \to p + e^- + \tilde \nu_e$.
Then we address the compatibility of $\Delta$-driven phase transitions and dUrca reactions
with compact stars EoS that obey the $2M_{\odot}$ constraint \cite{Antoniadis2013}.
Finally the relevance of thermodynamic instabilities
in hot stellar matter beyond $\beta$-equilibrium is studied
by plotting the phase diagram.

\subsection{$U_{\Delta}^{(N)}$ potentials}
\label{ssec:UDN}

\begin{figure}[tb]
\centering
\includegraphics[width = 0.999\hsize]{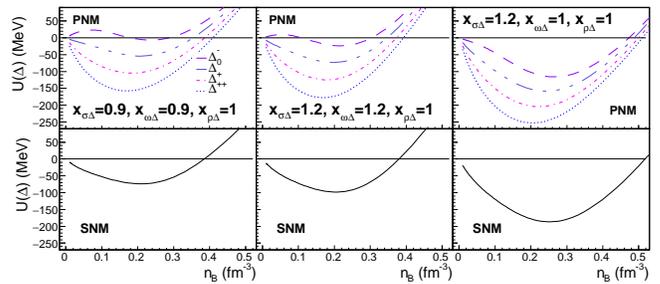}
\caption{Single-particle potentials of $\Delta$ isobars as a function of baryonic particle number density
  in symmetric nuclear matter (SNM) (bottom) and pure neutron matter (PNM) (top panels)
  for different values of $x_{\sigma \Delta}$ and  $x_{\omega \Delta}$, as indicated in each panel.
  In all cases $x_{\rho \Delta}=1$.
}
\label{fig:UDNM}
\end{figure}

Nucleation of $\Delta$-isobars obviously depends on their interaction
potential in nuclear matter,
\begin{equation}
  U_{\Delta}^{(N)}=-g_{\sigma \Delta} \bar \sigma+g_{\omega \Delta}
  \bar \omega + g_{\rho \Delta} t_{3} \bar \rho,
  \label{eq:UDN}
  \end{equation}
where $\bar \sigma$, $\bar \omega$, $\bar \rho$ stand for the mean field
expectation values of the mesonic field and $t_3$ represents the third
component of the isospin, with the convention $t_{3\Delta^{++}}=3/2$;
the dependence on baryonic particle number
densities has been omitted for the potential, coupling constants and
mesonic fields.
Eq. (\ref{eq:UDN}) suggests that, for a given nucleonic EoS,
more attractive potentials are provided by large values of $x_{\sigma \Delta}$
and low values of $x_{\omega \Delta}$.
It also shows that in neutron rich matter, as it is the case of NS,
the potential felt by positively charged isobars is larger in absolute values
than that felt by negative isobars since $\bar \rho<0$. At variance with this, in
symmetric nuclear matter (SNM), all isobars experience the same potential.

Quantitative information is provided in Fig. \ref{fig:UDNM} for the extreme
cases of pure neutron matter (PNM) and SNM.
Different combinations of $(x_{\sigma \Delta},x_{\omega \Delta})$ are considered;
for all cases we assume $x_{\rho \Delta}=1$.
It comes out that, depending on the coupling constants,
$U_{\Delta}^{(N)}$ in SNM is attractive for $0 <n_B \lesssim 2.5 - 5 n_s$.
For the PNM case a stronger dependence on the values of the coupling constants
is obtained along with a significant dispersion among the
predictions corresponding to different isobars.

\subsection{$\Delta$s in cold catalyzed matter}
\label{ssec:onset}

\begin{figure}[tb]
\centering
\includegraphics[width = 0.999\hsize]{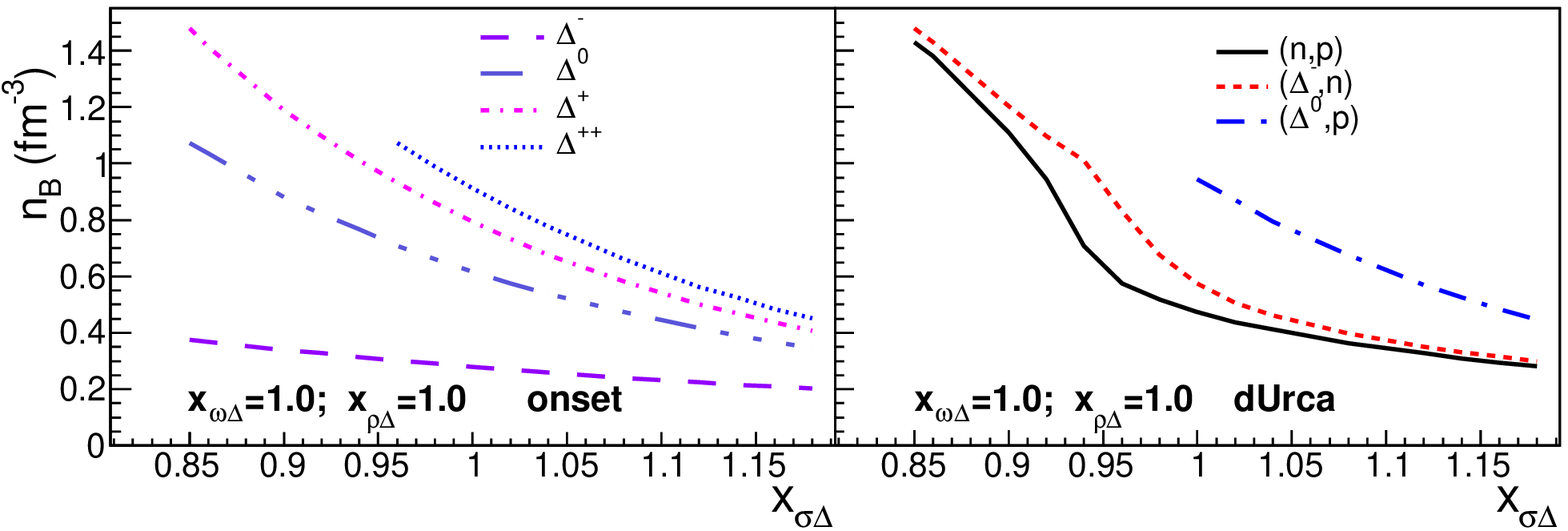}
\includegraphics[width = 0.999\hsize]{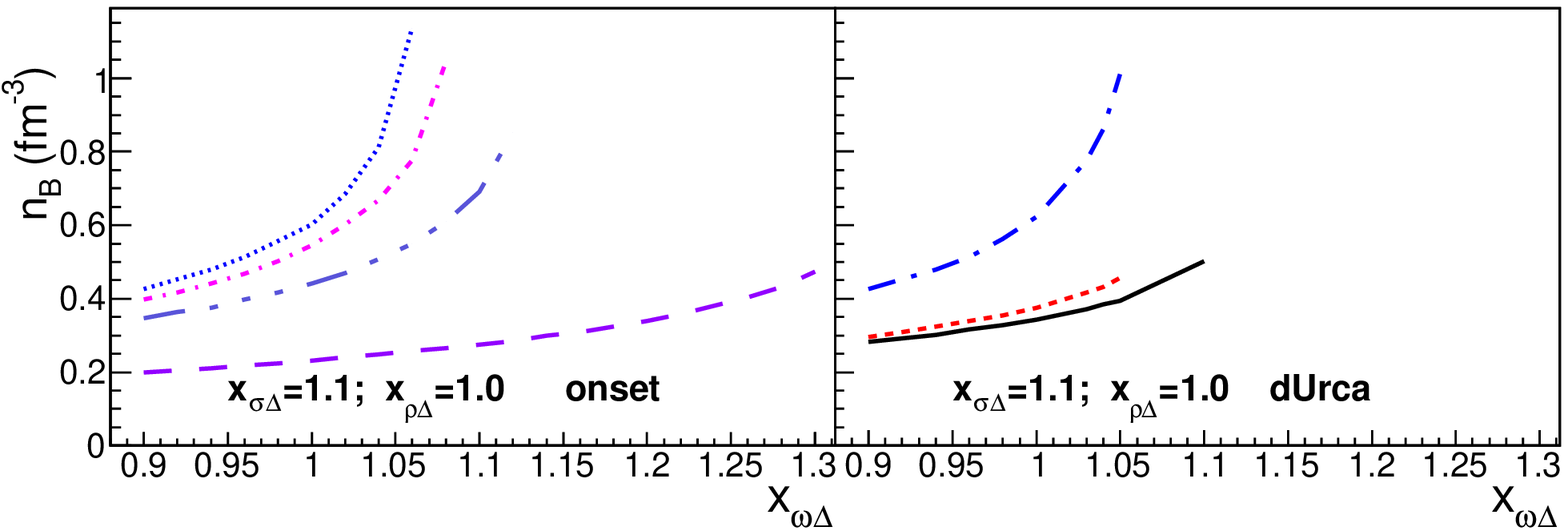}
\includegraphics[width = 0.999\hsize]{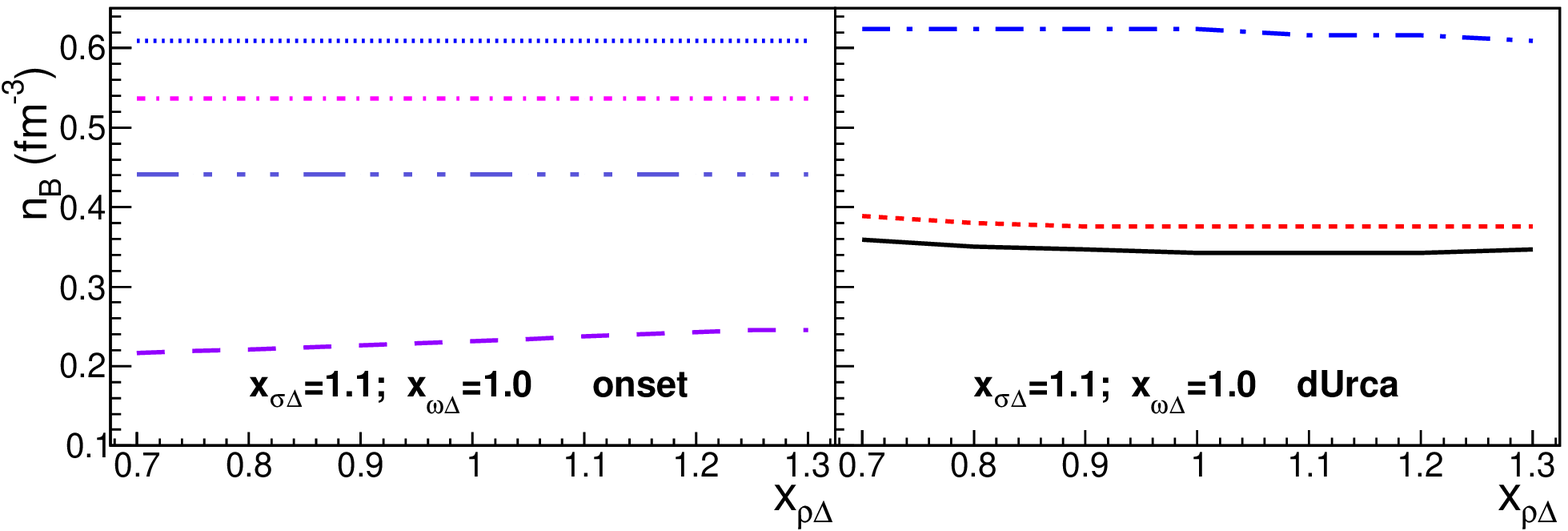}
\caption{Onset densities of $\Delta$ isobars (left)
  and threshold densities of various dUrca processes (right)
  in cold catalyzed neutron star (NS) matter
  as a function of $x_{\sigma \Delta}$ (top), $x_{\omega \Delta}$ (middle)
  and $x_{\rho \Delta}$ (bottom).
}
\label{fig:onset}
\end{figure}

The different attractive potentials felt by the different isobars in
neutron rich matter together with the chemical equilibrium conditions
\begin{equation}
  \mu_{\Delta^-}=\mu_B-\mu_Q; \mu_{\Delta^0}=\mu_B; \mu_{\Delta^+}=\mu_B+\mu_Q;
  \mu_{\Delta^{++}}=\mu_B+2\mu_Q,
  \label{eq:mueq}
\end{equation}
determine the onset densities of each species.
Here $\mu_B$ and $\mu_Q$ stand for baryon and charge chemical potentials and
can be expressed in terms of chemical potentials of nucleons, leptons and electrons as
$\mu_B=\mu_n$ and $\mu_Q=\mu_p-\mu_n=\mu_L-\mu_e$.
The left panels of Fig. \ref{fig:onset} illustrate the role played
by each coupling constant when the other two are kept constant.
For all considered sets of
($x_{\sigma \Delta}$, $x_{\omega \Delta}$, $x_{\rho \Delta}$)
the first particle that nucleates is $\Delta^-$.
Despite the low attractive potential its population is favored by charge
conservation.
The other three particles nucleate only for certain values of
the coupling constants; the second most favored particle is
$\Delta^0$, while the less favored one is $\Delta^{++}$.
We also note that:
i) the onset densities decrease (increase) with the increase of
$x_{\sigma \Delta}$ ($x_{\omega \Delta}$), reflecting thus the evolution of $U_{\Delta}^{(N)}$ with
the two coupling constants;
ii) $x_{\rho \Delta}$ has almost no influence on the onset densities;
iii) the onset density of $\Delta^-$ is significantly lower than those
of the other three particles, which do not differ much one from the other.

By partially replacing the electrons, which regulate the relative abundances of
neutrons and protons, $\Delta^-$ will implicitly modify the neutron and
proton abundances. If protons will become sufficiently abundant
such as the triangle inequalities of Fermi momenta are satisfied \cite{DU91}, nucleonic
dUrca - forbidden in DDME2 - will become energetically allowed.
If, additionally, $\Delta$s
are abundant enough to satisfy, along with electrons and nucleons,
the appropriate triangle inequalities, also dUrca
processes involving these particles will open up.

The right panels of Fig. \ref{fig:onset} illustrate the threshold densities of
nucleonic dUrca as well as those of two dUrca processes involving $\Delta$s
as a function coupling constants. The same cases as in the left panels
are considered. With the exception of $x_{\sigma \Delta}=1.1$ and
$x_{\omega \Delta} \geq 1.1$, all considered sets of parameters allow
for nucleonic dUrca, otherwise forbidden in NS built upon DDME2.
$\Delta^- \to n + e^- + \tilde \nu_e$ is energetically allowed
roughly over the same parameter ranges which allow for nucleonic dUrca;
its threshold density is only slightly higher than the one of nucleonic dUrca.
$\Delta^0 \to p + e^- + \tilde \nu_e$ opens up at higher densities and
operates over a reduced range of parameters.
Similarly to the onset densities, density thresholds of dUrca processes
show strong (poor) sensitivity to $x_{\sigma \Delta}$ and $x_{\omega \Delta}$
($x_{\rho \Delta}$).

\subsection{Spinodal instabilities in cold catalyzed matter}
\label{ssec:instab}

\begin{figure}[tb]
\centering
\includegraphics[width = 0.9\hsize]{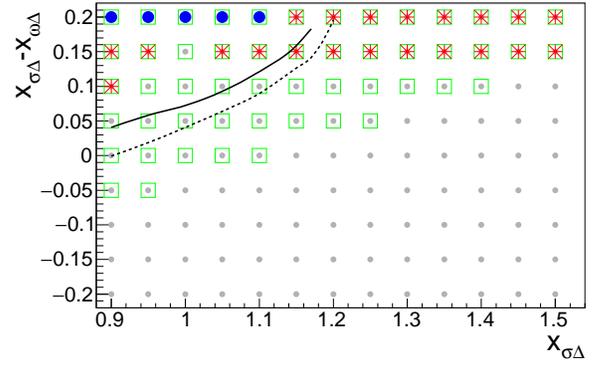}
\caption{Range of coupling constants $(x_{\sigma \Delta},x_{\omega \Delta})$
  for which cold $\beta$-stable matter features one (red stars) or two (blue dots)
  spinodal instability domains.
  Green squares: EoS which allow for at least one dUrca process at $n_B<0.8$ fm$^{-3}$.  
  Solid (dashed) line: contours of coupling constants corresponding to
  $M_{\rm max}=2.0M_{\odot}$ ($2.2M_{\odot}$). Results corresponding to $x_{\rho \Delta}=1$.
}
\label{fig:phasetrans}
\end{figure}

\begin{figure*}[tb]
\centering
\includegraphics[width = 0.45\hsize]{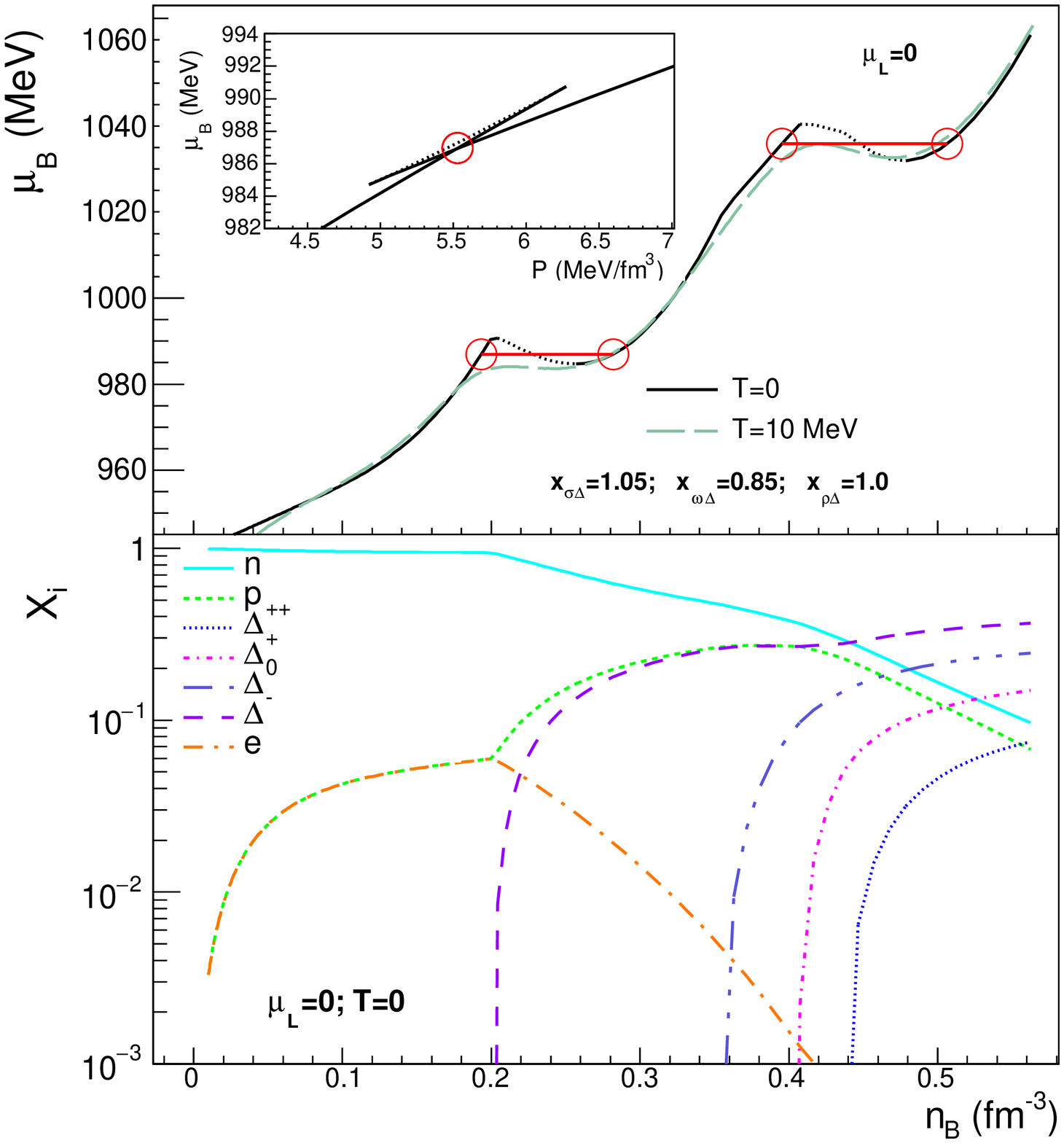}
\includegraphics[width = 0.45\hsize]{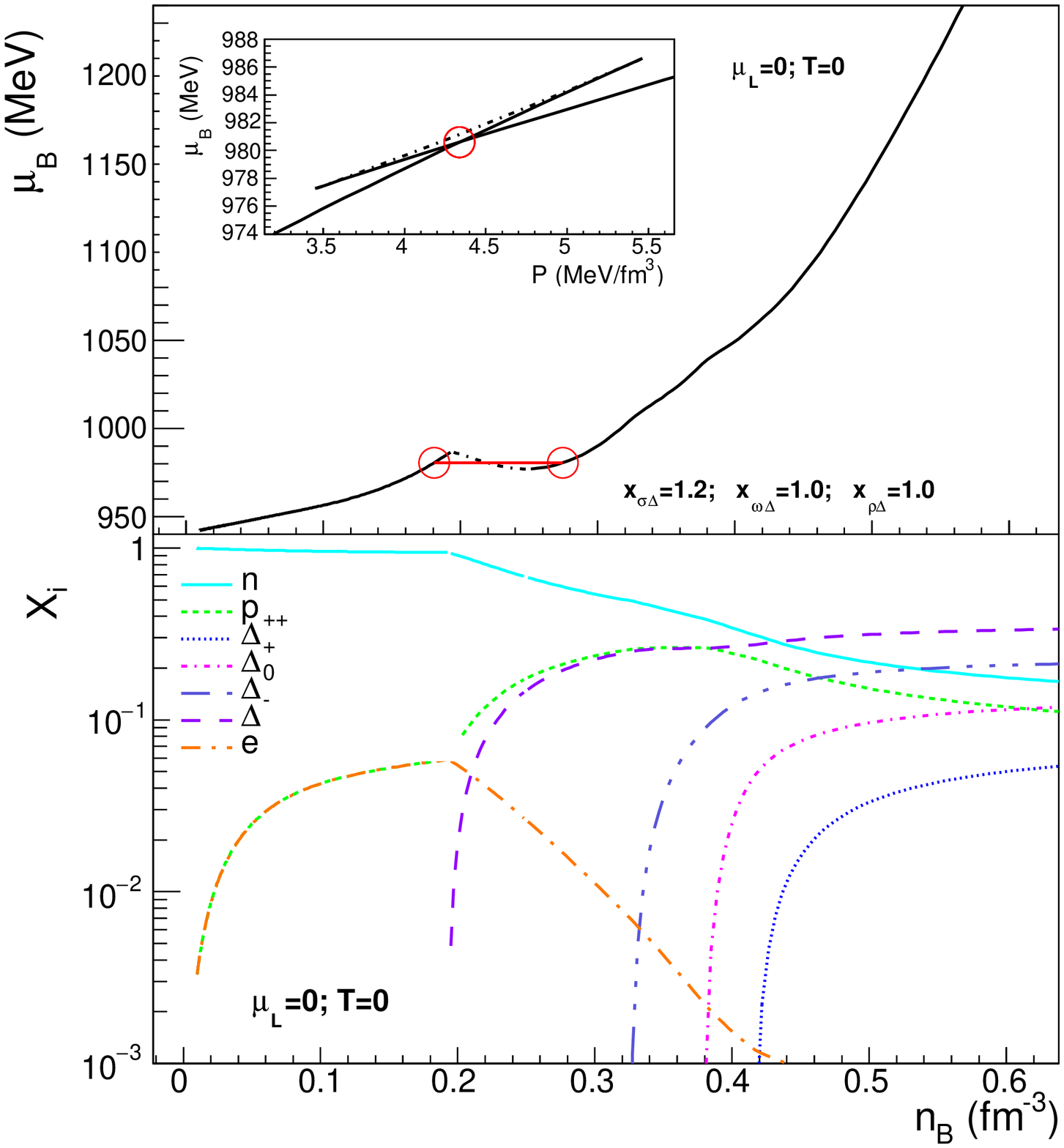}
\caption{Baryonic chemical potential (top) and relative particle densities (bottom)
  as a function of baryonic particle number density
  for $\Delta$-admixed 
  $\beta$-equilibrated neutrino-transparent matter at $T=0$.
  The insets in the top panels depict the baryonic chemical potential as a function of total pressure
  over a density domain which contains the threshold for the onset of $\Delta^-$s.
  Red open circles and horizontal segments indicate the Gibbs constructions.
  Stable (unstable) solutions of mean-field calculations are illustrated
  with solid (dotted) lines.  
  Results corresponding to $x_{\sigma \Delta}=1.05$, $x_{\omega \Delta}=0.85$ (left)
  and $x_{\sigma \Delta}=1.20$, $x_{\omega \Delta}=1.00$ (right) and
  $x_{\rho \Delta}=1.00$. 
  On the top left panel also shown is $\mu_B (n_B)$ for $T$=10 MeV.
}
\label{fig:muBnB}
\end{figure*}

We now turn to investigate the occurrence of $\Delta$-driven instabilities
in cold catalyzed matter. With the purpose to limit the parameter space
and the motivation that $x_{\rho \Delta}$ does not impact the onset of $\Delta$s,
we fix $x_{\rho \Delta}=1$.

We generate EoS of $\Delta$-admixed cold $\beta$-equilibrated matter in the
neutrino-transparent regime
for $0.9 \leq x_{\sigma \Delta} \leq 1.5$
and $x_{\sigma \Delta}-0.2 \leq x_{\omega \Delta} \leq x_{\sigma \Delta}+0.2$.
Thermodynamic instabilities are sought after by calculating the eigenvalues
of the curvature matrix $C_{B;L}=\partial^2 e(n_B,n_L)/\partial n_B \partial n_L$.
Three situations are encountered:
a) no instability,
b) one domain of instabilities,
c) two domains of instabilities.
The sets of $\left( x_{\sigma \Delta}, x_{\omega \Delta}\right)$ corresponding
to the latter two situations are illustrated in Fig. \ref{fig:phasetrans}
with red stars and, respectively, blue solid dots.
Instabilities occur for a narrow domain of
$0.15 \leq x_{\sigma \Delta}-x_{\omega \Delta} \leq 0.2$ and,
for the lowest values of $x_{\sigma \Delta}$, $0.9 \leq x_{\sigma \Delta} \leq 1.1$,
two instability domains are present. For all matter states featuring spinodal instability
only one negative eigenvalue was found for $C_{B;L}$.

Further insight into the thermodynamic behavior
is provided in Fig. \ref{fig:muBnB} in terms
of $\mu_B(n_B)$ (top panels); the chemical composition as a function
of baryonic particle number density is represented as well (bottom panels).
Two sets $\left( x_{\sigma \Delta}, x_{\omega \Delta}\right)$
are considered, as mentioned in left and right panels.
For the lower value of $x_{\sigma \Delta}$ two instability domains
are obtained, signaled by backbendings of $\mu_B(n_B)$.
Inspection of the bottom panel indicates that the first instability is
due to the onset of $\Delta^-$ and the second one to the onset of the other
three isobars.
For the higher value of $x_{\sigma \Delta}$ only the low density instability
related to nucleation of $\Delta^-$ is present.
The fact that - despite their earlier onset - $\Delta^0$, $\Delta^+$ and
$\Delta^{++}$ do not lead to instabilities is due to the less steep
increase of their abundances, caused by the higher value of $x_{\omega \Delta}$.

$\mu_B(n_B, \mu_L=0)$ considered in Fig. \ref{fig:muBnB} corresponds
to hybrid ensembles which allow multi-component systems be treated
as uni-component ones and have the Gibbs construction reduced to the
Maxwell construction \cite{Ducoin_NPA_2006}.
Coexisting phases, characterized by equal values of
intensive quantities, {\em i.e.} pressure ($P$) and the two chemical
potentials ($\mu_B$ and $\mu_L$), are represented with red open dots. Complementary
information is provided in the insets in terms of $P(\mu_B)$.

The instabilities induced by $\Delta^-$ occur over
$n_s \lesssim n_B \lesssim 2n_s$, which corresponds to the outer core of NS.
As such it is expected to impact the radii of all mass NS and
have little or no effect on the maximum mass.
The instabilities induced by $\left(\Delta^0, \Delta^+, \Delta^{++} \right)$
occur over $2.5 n_s \lesssim n_B \lesssim 4 n_s$, which corresponds to
more central shells in the core.
In addition to radii also the maximum NS mass may be affected.
However some of the coupling constant sets might not be
compatible with the $2M_{\odot}$ constraint \cite{Antoniadis2013}.
To check the issue we plot in Fig. \ref{fig:phasetrans} the
sets of $\left( x_{\sigma \Delta}, x_{\omega \Delta}\right)$
for which the maximum NS mass equals $2M_{\odot}$ and $2.2M_{\odot}$.
Only the coupling constants sets lying at the r.h.s. of these
curves provide maximum NS masses larger than the value to which the
contour corresponds. This means that the occurrence of thermodynamic
instabilities associated to $\left(\Delta^0, \Delta^+, \Delta^{++} \right)$
is ruled out by the astrophysical limit on pulsar masses.
This is the case of EoS discussed by \cite{Lavagno_2019} which,
indeed, provide low mass NS.
The explanation of the low maximum NS masses obtained for this domain of the parameter space
consists in the fact that, because of vanishing nucleon effective masses, the baryonic
density can not exceed $3-4 n_s$.
In conclusion the only phase transition compatible with
$2M_{\odot}$ is the low density one.

Also shown in Fig. \ref{fig:phasetrans} is the domain
of parameter sets for which the density threshold of the first
energetically allowed dUrca process is smaller than
0.8~fm$^{-3}$. This arbitrarily chosen value roughly coincide
with the central baryonic density of the maximum mass configuration provided
by DDME2 when only nucleonic d.o.f. are accounted for.

\subsection{The phase diagram of $(n,p,\Delta,e)$ matter}
\label{ssec:phdiagram}

\begin{figure}[tb]
\centering
\includegraphics[width = 0.9\hsize]{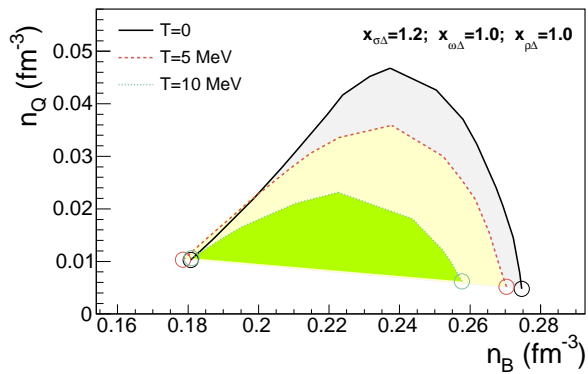}
\caption{Phase coexistence domains (shaded regions) between pure $(n, p, e)$-matter and
  $\Delta$ admixed matter in $n_B-n_Q$ coordinates for $T=0$, 5 and 10 MeV.
  Open circles correspond to coexisting phases at $\mu_L=0$.
  Results corresponding to $x_{\sigma \Delta}=1.2$, $x_{\omega \Delta}=1.0$, $x_{\rho \Delta}=1.0$.
}
\label{fig:phasediagram}
\end{figure}

The phase diagram of $(n,p,\Delta,e)$ matter corresponding to the coupling
constants set considered in the right panels of Fig. \ref{fig:muBnB}
is plotted in Fig. \ref{fig:phasediagram}. In addition to $T=0$ also the domains
of phase coexistence between pure $(n,p,e)$-matter and $\Delta$-resonance admixed matter
for $T=5$ and 10 MeV are represented.
As in Fig. \ref{fig:muBnB} the open dots mark, for each temperature and $\mu_L=0$,
the coexisting phases.
It comes out that, contrary to the liquid-gas phase transition taking place in
sub-saturated nuclear matter, the order parameter of the $\Delta$-driven phase transition
has only a tiny component along $n_Q$.
Other features are nevertheless similar to those of the liquid-gas phase transition: 
the coexistence domain gets quenched as the temperature increases and second
order phase transitions occur as well.
First order phase transitions in hot and dense matter have been shown to impact
star stability with respect to collapse to a black hole during the core collapse \cite{Peres_PRD_2013}.
Second order phase transitions have been proven to drastically
reduce the neutrino mean free path and, thus, slow down the proto-NS cooling \cite{Gulminelli_PRC_2013}.
Though addressed in connection with a strangeness driven phase transition associated to hyperonisation
the two effects are expected to hold also in the case of the presently discussed phase transition
and, thus, lead to observable effects.

\subsection{Global properties of NS featuring a $\Delta$-driven phase transition}
\label{ssec:MR}

\begin{figure}[tb]
\centering
\includegraphics[width = 0.9\hsize]{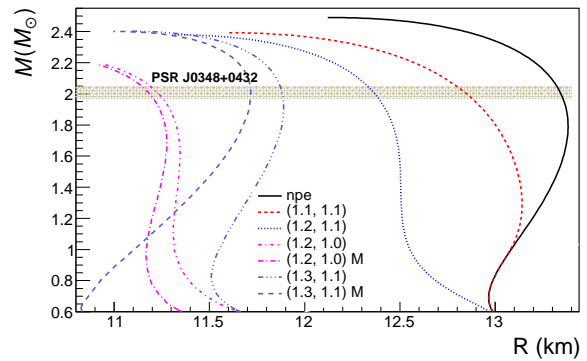}
\caption{Gravitational mass - radius diagram of cold catalyzed NS built upon DDME2.
  Results corresponding to $\Delta$-admixed matter are confronted with those
  corresponding to purely nucleonic matter for different values of the
  coupling constants of mesons to $\Delta$-isobars, specified as
  ($x_{\sigma \Delta}$, $x_{\omega \Delta}$). In all cases $x_{\rho \Delta}=1$.
  EoS featuring a phase transition where the Maxwell construction
  was performed are marked with "M".
  The shaded horizontal stripe shows the observed mass of pulsars PSR J0348+0432,
  $M= 2.01 \pm 0.04 M_{\odot}$ \cite{Antoniadis2013}.
}
\label{fig:MR}
\end{figure}

\begin{figure}[tb]
\centering
\includegraphics[width = 0.9\hsize]{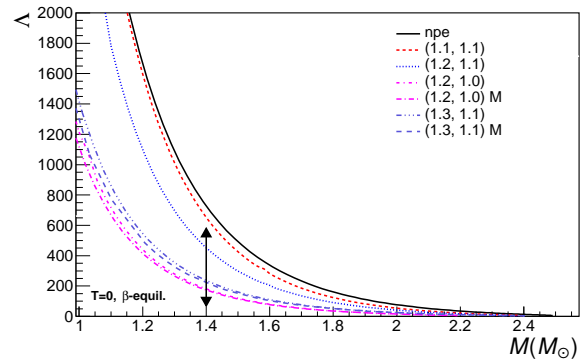}
\caption{Tidal deformabilities as a function of gravitational mass for
  the same EoS as in Fig. \ref{fig:MR}.
  The vertical arrow marks the limits obtained for a $1.4M_{\odot}$ NS,
  $70 < \Lambda_{1.4} < 580$, as derived from the observation of GW170817 by the LVC
  collaboration \cite{Abbott_PRL_18}.
}
\label{fig:LM}
\end{figure}

In the following we discuss some properties of $\Delta$-admixed NS
with focus on modifications induced by the phase transition.
Equilibrium configurations of spherically-symmetric non-rotating NS
are obtained by solving the Tolman-Oppenheimer-Volkoff set
of differential equations. The tidal deformabilities are calculated following
\cite{Hinderer_ApJ_2008,Hinderer_PRD_2010}.

Figs. \ref{fig:MR} and \ref{fig:LM} illustrate the mass-radius diagram
and, respectively, the tidal deformabilities as a function of gravitational mass.
Results corresponding to several $\Delta$-admixed NS are confronted with those of purely nucleonic stars.
As previously discussed, nucleation of $\Delta$s entails a reduction of the maximum mass
of up to a few tens of solar mass \cite{Cai_PRC_2015,Zhu_PRC_2016}
and a diminish of NS radii by up to 1-2 km~\cite{Drago_PRC_2014,Kolomeitsev_NPA_2017,Li_PLB_2018,Ribes_2019,Li_ApJL_2019}. Related to the latter effect a significant reduction of tidal deformabilities is also obtained.
The amplitude of these modifications increases with the amount of heavy baryons, that is with
the increase of $x_{\sigma \Delta}$ and the decrease of $x_{\omega \Delta}$.
For the two considered sets of coupling constants for which matter manifests instabilities
the Maxwell construction causes an extra reduction of the order of several percent
of both radii and tidal deformabilities.

\section{Conclusions}
\label{sec:concl}

Inspired by previous findings of Lavagno and Pigato \cite{Lavagno_2019} regarding spinodal instabilities
in cold catalyzed NS matter with admixture of $\Delta(1232)$-resonances we have investigated
in a systematic way the parameter space of the $\Delta-N$ interaction and identified the domains
where the onset of $\Delta$ generates thermodynamic instabilities.
EoS which fulfill the two solar mass astrophysics constraint manifest spinodal instabilities over
$n_s \lesssim n_B \lesssim 2 n_s$, which corresponds to the outer shells of the core.
When mean-field anomalous behavior is cured by Maxwell construction an extra diminish,
of the order of several percent,
is obtained for both NS radii and tidal deformabilities.
$\Delta^-$-driven instabilities persist also out of $\beta$-equilibrium in the
neutrino-transparent matter and at temperatures of a few tens MeV,
which suggests that - if present - the thermodynamic instabilities may affect
also other astrophysical phenomena where dense matter is populated.

Finally we have shown that, for a large domain of the parameter space, nucleation of $\Delta$s
opens-up the nucleonic dUrca process which is otherwise forbidden.
The consequences on thermal evolution of isolated NS and X-ray transients are beyond the aim
of this work and will addressed elsewhere.
Nevertheless it is straightforward to anticipate that $\Delta$-admixed massive NS,
whose inner core accommodates unpaired nucleons, might meet the conditions required to describe
the data of the faintest X-ray transients,
XRT SAX J1808.4-3658 and 1H 1905+000
\cite{BY15}.
The situation is interesting the more as to date none of the covariant density functional models
with coupling constants fitted on experimental and observational data has been able to simultaneously describe
the whole set of thermal data from isolated NS and accreting NS in quiescence,
even if the pairing gaps in different channels were allowed to
vary over wide ranges and hyperonic degrees of freedom have been accounted for \cite{Fortin_cooling}.

\section*{Acknowledgements}
This work was, in part, supported by the European COST Action "PHAROS"
(CA16214). The author acknowledges support from UEFISCDI via grant nr.
PN-III-P4-ID-PCE-2020-0293.

\end{document}